\documentclass[fleqn,usenatbib]{mnras}

\usepackage[T1]{fontenc}

\DeclareRobustCommand{\VAN}[3]{#2}
\let\VANthebibliography\thebibliography
\def\thebibliography{\DeclareRobustCommand{\VAN}[3]{##3}\VANthebibliography}

\newcommand{\teff}{$T_\mathrm{eff}$}
\newcommand{\logg}{$\log g$}
\newcommand{\logy}{$\log n(\mathrm{He})/n(\mathrm{H})$}

\usepackage{graphicx}	% Including figure files
\usepackage{amsmath}	% Advanced maths commands
\usepackage{amssymb}	% Extra maths symbols
\usepackage{caption}
\usepackage{subcaption}
\usepackage{setspace}
\usepackage{newtxtext,newtxmath}

\title[SALT hydrogen-deficient binaries]{A search for close binary systems in the SALT survey of hydrogen-deficient stars using {\it TESS}}

\author[E. J. Snowdon et al.]{
E. J. Snowdon$^{1}$\thanks{E-mail: edward.snowdon@armagh.ac.uk},
C. S. Jeffery$^{1}$,
S. Schlagenhauf$^{1,2}$,
M. Dorsch$^{3,4}$
\\
$^{1}$Armagh Observatory and Planetarium, College Hill, Armagh, BT61 9DB, UK\\
$^{2}$School of Mathematics and Physics, The Queen's University of Belfast, University Road, BT7 1NN, UK\\
$^{3}$Institut für Physik und Astronomie, Universität Potsdam, Haus 28, Karl-Liebknecht-Str. 24/25, 14476 Potsdam-Golm, Germany\\
$^{4}$Dr. Karl Remeis-Observatory \& ECAP, Friedrich-Alexander University Erlangen-Nürnberg, Sternwartstr. 7, 96049 Bamberg, Germany\\
}

\date{Accepted XXX. Received YYY; in original form ZZZ}

\pubyear{2015}

\begin{document}
\label{firstpage}
\pagerange{\pageref{firstpage}--\pageref{lastpage}}
\maketitle

\begin{abstract}
The {\it TESS} periodograms of the SALT survey catalogue of hydrogen-deficient stars were searched for evidence of short-period variability.
Periodic light curve variations were identified in 16 stars out of 153 catalogue objects, of which 10 were false positives. 
From the remaining 6 identified variables, Ton S 415 is a known close binary system and the sixth close binary containing a hydrogen-deficient hot subdwarf.  
Radial velocity and SED analyses ruled out the remaining 5 as close binary systems; the causes of their variability remain uncertain. 
With one or more K-type  companions, BPS CS 22956-0094 may be a wide binary or triple. 
From this SALT+{\it TESS} sample, the fraction of close binaries stands at $ 1/29 \approx 3.5\%$ for intermediate helium hot subdwarfs and $0/124 = 0\%$ for extreme helium subdwarfs. 
\end{abstract}


%%%%%%%%%%%%%%%%%%%%%%%%%%%%%%%%%%%%%%%%%%%%%%%%%%

%%%%%%%%%%%%%%%%% BODY OF PAPER %%%%%%%%%%%%%%%%%%

\section{Introduction}

The hot subdwarfs are a diverse group of low-mass, subluminous stars that lie between the main sequence and the white dwarfs on the Hertzsprung-Russell diagram, in a region known as the extended horizontal branch \citep{heber86}. They can be categorised according to their atmospheric helium compositions; most have helium-deficient atmospheres with abundances $\leq1\%$ by number, as radiative levitation and atomic diffusion processes cause helium to settle below the hydrogen-rich surface layers \citep{heber84}. A minority of around $10\%$ of hot subdwarfs instead have hydrogen-deficient atmospheres dominated by strong helium lines, with corresponding helium abundances of $20-99.9\%$ \citep{green86}.

The exact processes by which hot subdwarfs form and evolve remain uncertain. The diverse and interconnected nature of the extended horizontal branch stars means multiple evolutionary models are required to account for different populations. The most widely-accepted formation channels for hydrogen-rich hot subdwarfs were proposed by \citet{han02} and \citet{han03}, in which a red giant star in a binary system overfills its Roche lobe while ascending the RGB. If the resulting mass transfer is unstable, a common envelope forms around both components. The orbital separation shrinks as energy is lost to friction until the envelope is ejected, leaving the degenerate core of the giant progenitor (which subsequently ignites helium to become a hot subdwarf) and the companion in a close binary ($p\leq10\,{\rm d}$). Evolutionary models for hydrogen-deficient hot subdwarfs are commonly based on mergers. \citet{han03} proposed that helium-rich B subdwarfs (He-sdB) may form from the merger of two helium white dwarfs. Helium-rich O subdwarfs (He-sdO) have been proposed to form from the merger of He-core white dwarfs, producing extreme helium stars which then contract towards the helium main sequence \citep{iben90, saio00}.

The Roche lobe overflow models are supported by the high fraction ($\sim60\%$) of sdBs found in close binaries \citep{maxted01}.  \citet{han03} predicted an intrinsic fraction of $76-89\%$, but anticipated that observed fractions would be lower than this due to selection effects. Conversely, the merger models imply that hydrogen-deficient subdwarfs should not exist in close binaries. However, examples of hydrogen-deficient binaries have been found; \citet{ahmad04} identified the star PG 1544$+$488 to be a double He-sdB binary, while \citet{stroeer07} found evidence for a double He-sdO binary. These discoveries challenge the current understanding of hot subdwarf evolution, but the rarity and diversity of the known hydrogen-deficient binaries make it difficult to establish a basis for new evolutionary models.

Since 2016, multiple observing programmes at the Southern African Large Telescope (SALT) have focused on analyses of hydrogen-deficient stars. The first outcome of this survey was to use intermediate- and high-resolution spectroscopy to classify all programme stars according to the scheme of \citet{drilling13}, as well as to determine their atmospheric parameters and chemical abundances. These results are being used to identify chemically-peculiar stars, to test evolutionary models, to search for links between populations, and to identify hydrogen-deficient binaries. Many stars from the SALT catalogue have also been observed by the Transiting Exoplanet Survey Satellite ({\it TESS}), allowing binaries to also be identified from photometric variability \citep{tess}. This paper presents the results of a search for close binary candidates from the SALT catalogue of hydrogen-deficient stars, using both SALT spectra and {\it TESS} light curves.

\section{Observations}
\begin{table*}
    \setlength{\tabcolsep}{3pt}
    \centering
    \begin{tabular}{lllccccccl}
        \hline
        {\it Gaia} DR3 & TIC & Name & $S$ & SAP & $p$ (d) & $a$  & Class & $m_{\rm G}$ & Notes \\%& Source \\
        \hline
        \multicolumn{10}{c}{\it Identified Variables}\\ 
        6404530097924835968 & 234281664 & BPS CS 22956$-$0094 & 3 & 0.9948 & 0.388 & 0.021 & sdB0.5V:He24 & 13.3 & \\%& [2] \\
        3176695152292552704 & 332697630 & EC 04110$-$1348 & 2 & 0.9960 & 0.488 & 0.005 & sdOC7.5VII:He39 & 12.5 & \\%& [1] \\
        6479757549625889536 & 79246796 & EC 21077$-$4815 & 2 & 0.7903 & 0.815 & 0.015 & sdOC7.5VII:He39 & 15.2 & \\%& [2] \\
        4962221462214625920 & 115273584 & SB 705 & 2 & 0.9855 & 0.505 & 0.005 & sdOC7.5VII:He40 & 12.9 & \\%& [3] \\
        4827575749313476096 & 77959296 & Ton S 415 & 2 & 0.9918 & 0.058 & 0.003 & sdO8VII:He30 & 13.3 & sdOB-WD binary\\%& [1] \\
        6158218941883819648 & 130945039 & TYC 7242$-$541$-$1 & 2 & 0.9370 & 0.396 & 0.010 & sdOC6VI:He39 & 12.3 & \\
        \multicolumn{10}{c}{\it Rejected Candidates}\\ 
        6693254326596860288 & 290237162 & BPS CS 22885$-$0043 & 1 & 0.8423 & 0.359 & 0.007 & sdOC6.5VI:He39 & 15.8 & Bad aperture mask\\%& [1] \\ 
        3073231760953563264 & 121318590 & [CW83] 0832$-$01 & 3 & 0.9429 & 1.85 & 0.001 & sdO8VII:He40 & 11.4 & Contaminated \\%& [4] \\
        2978785258317207168 & 121641254 & EC 04399$-$1826 & 1 & 0.9736 & 1.486 & 0.002 & sdO8VI:He39 & 14.2 & Poor S/N \\%& [1] \\
        6700716918173536768 & 48939456 & EC 20081$-$3205 & 1 & 0.5748 & 0.342 & 0.125 & sdON5VII:He40 & 16.2 & Contaminated \\%& [1] \\
        6458088786782625024 & 219304041 & EC 20577$-$5504 & 1 & 0.4607 & 0.254 & 0.018 & sdO9VI:He37 & 15.8 & Bad aperture mask\\%& [1] \\
        6512436478313050496 & 161358962 & EC 22536$-$5304 & 1 & 0.8123 & 0.026 & 0.005 & sdB0.2VI:He23 & 13.24 & Crowded frame\\
        5747322001951403776 & 289533626 & GALEX J084528.7$-$12140 & 3 & 0.7935 & 0.702 & 0.002 & sdOC9.5VI:He39 & 14.0 & Crowded frame \\%& [5] \\
        6095832381546853888 & 242416400 & HD 124448 & 3 & 0.9078 & 0.819 & 0.004 & sdBCIII:He40 & 9.94 & Known pulsator\\
        2724882872133120512 & 257044138 & PG 2158$+$082 & 1 & 0.8183 & 0.635 & 0.005 & sdO2VII:He40 & 13.1 & Crowded frame \\
        5926622760411844992 & 213654615 & UCAC4 204$-$139111 & 2 & 0.2265 & 0.242 & 0.010 & sdO2VI:He38 & 13.5 & Contaminated \\
        \hline
    \end{tabular}
    \caption{SALT survey objects with periods in their {\it TESS} periodograms detectable above the $8\sigma$ threshold. 
    Objects have been divided into confirmed variables and rejects; see text for more information.  
    $S$ denotes the number of {\it TESS} sectors in which observations for a star are available. Amplitude values ($a$) are the peak-to-peak semi-amplitude of the normalised light curve. CROWDSAP values are taken from the mean of all sectors. Spectral classifications on the scheme of \citet{drilling13} are from the ongoing SALT survey \citep{jeffery20}. G-band apparent magnitudes taken from the {\it Gaia} mission \citep{gaia1, gaia2}. }
    \label{tab:tess_objects}
\end{table*}

\subsection{{\it TESS} data}

The original SALT survey was based on classifications of He-sdB, He-sdOB, and He-sdO stars taken from the Edinburgh-Cape survey \citep{stobie97, kilkenny97b}, with similar hydrogen-deficient stars being added from compilations such as \citet{beers92} and \citet{geier17}, among others. A complete description of this original SALT survey may be found in \citet{jeffery20}. The sample used for this work consisted of 664 objects, including those described by \citet{jeffery20} and others which have been observed subsequently. Candidate binaries were selected from the objects in the sample which had been observed in at least one {\it TESS} sector. {\it TESS} divides the sky into $24^{\circ}\times90^{\circ}$ sectors which are observed for 27 consecutive days at a time. The classification system of \citet{drilling13} assigns hot subdwarfs a helium class of between 0 (H-dominated) and 40 (He-dominated), denoting the strengths of He{\sc i} and{\sc ii} relative to the Balmer lines. To ensure only hydrogen-deficient objects were chosen, only objects with a helium class $\geq 15$ were considered. 395 of the 664 survey objects were found to be sufficiently hydrogen-deficient. These were then further classified as having either intermediate ($15\leq{\rm He}<35$) or extreme ($35\leq\rm{He}\leq40$) helium abundances. 291 objects were classified as extreme helium objects and 104 as intermediate. The light curves and periodograms of the qualifying stars were systematically searched for evidence of variability. 

Acquisition, manipulation, and analysis of light curve data was performed using the {\sc lightkurve} Python package \citep{lightkurve}. For a given target, all $120\,{\rm s}$ cadence data made available by the {\it TESS} Science Processing Operations Centre (SPOC) were downloaded from the Barbara A. Mikulski Archive for Space Telescopes (MAST) portal. Pre-search Data Conditioning Simple Aperture Photometry (PDCSAP) data were used to mitigate artefacting and instrumental effects \citep{kinemuchi12}. The light curves were flattened using a Savitzky-Golay filter with a polynomial order of 2 \citep{savitzky64} to correct any remaining low-frequency instrumental trends. 

\begin{figure}
    \centering
    \includegraphics[width=\columnwidth]{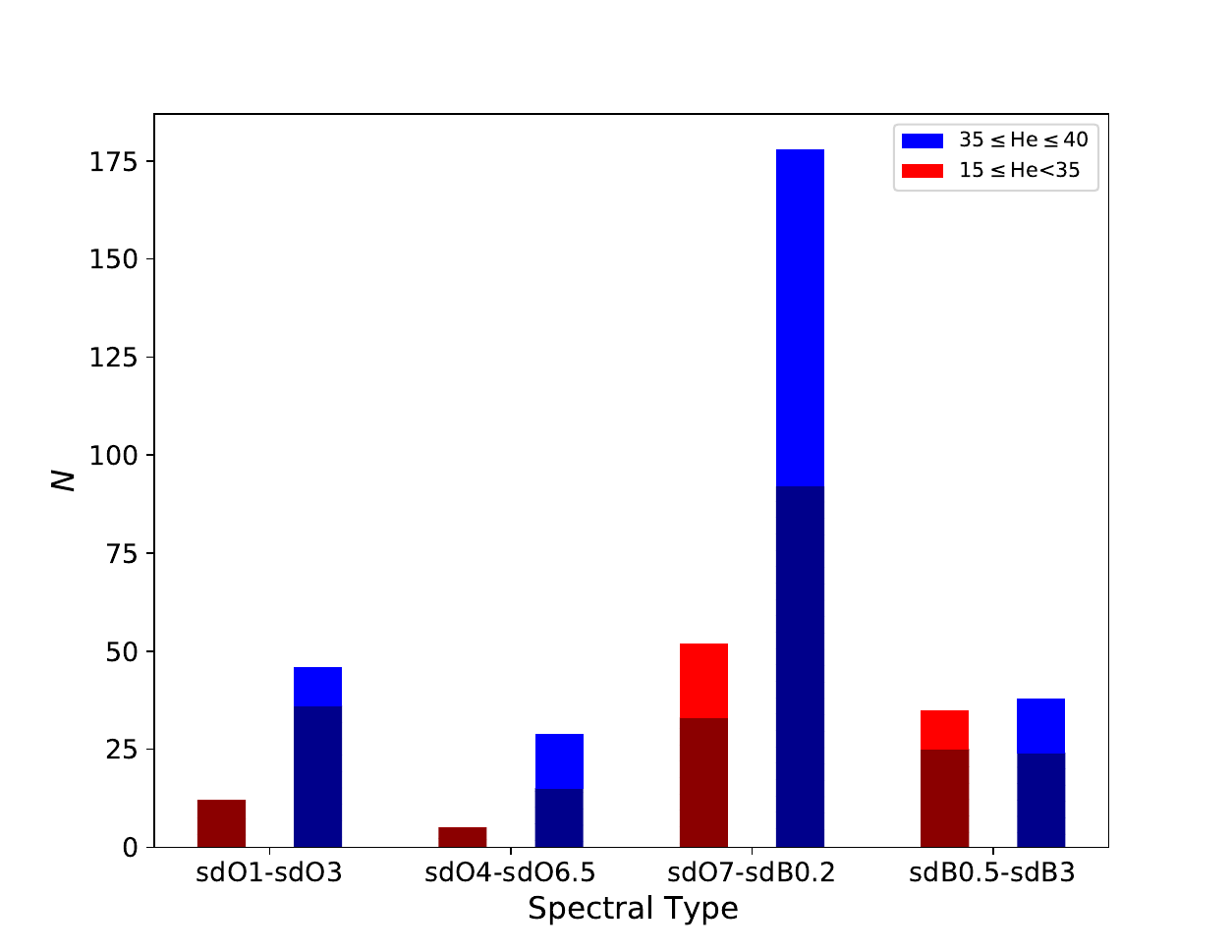}
    \caption{Classification summary for 395 hydrogen-deficient stars in the SALT survey based on Drilling spectral type. Red bars indicate stars with intermediate He abundances and blue bars indicate stars with extreme He abundances. The hatched portions of each bar represent objects without available 120\,s-cadence {\it TESS} data.}
    \label{fig:saltsurveyclasses}
\end{figure}

Out of 395 hydrogen-deficient objects in the sample, 120\,s-cadence {\it TESS} light curves were available for 153, of which 124 were extreme helium objects and 29 were intermediate. Of these, 16 objects were found with detectable periods. A breakdown of the sample by spectral type and helium class is shown in Fig.\ref{fig:saltsurveyclasses}. A summary of the identified candidates is given in Table \ref{tab:tess_objects}. A complete list of assessed target stars is given in Appendix \ref{app:tess}. The star Ton S 415 was singled out for detailed individual analysis due to its intermediate helium classification and very short orbital period \citep{snowdon23}.

\subsection{SALT data}

\begin{table}
    \centering
    \begin{tabular}{lcccc}
    \hline
    Target Name & Date & UT (start) & UT (end) & Exposures \\
    \hline
    BPS CS & 2023-05-03 & 03:50:59 & 03:54:18 & 2 \\ %90s
    22956$-$0094 & 2023-05-04 & 03:38:09 & 03:54:07 & 4 \\ %90/60s
     & 2023-05-05 & 03:03:07 & 03:22:45 & 2 \\ %90/20s
     & 2023-05-06 & 03:03:02 & 03:06:21 & 2 \\ %90s
     & 2023-05-07 & 03:05:32 & 03:20:28 & 4 \\ %90/60s
     & 2023-05-08 & 02:55:06 & 02:57:25 & 2 \\ %60s
     & 2023-05-09 & 02:45:55 & 02:48:15 & 2 \\ %60s
     \rule{0pt}{1ex} \\
    EC & 2020-08-01 & 03:28:08 & 03:40:57 & 4 \\
    04110$-$1348 & 2021-07-25 & 03:57:29 & 04:11:08 & 4 \\
     & 2021-07-31 & 03:55:38 & 04:09:16 & 4 \\
     & 2021-08-02 & 03:53:50 & 04:07:28 & 4 \\
     & 2021-08-07 & 03:44:06 & 03:57:44 & 4 \\
     & 2021-08-23 & 02:15:45 & 02:29:25 & 4 \\
     & 2021-08-25 & 02:10:59 & 02:24:37 & 4 \\
     & 2021-11-14 & 20:34:36 & 20:48:15 & 4 \\
     & 2022-01-02 & 21:56:58 & 22:10:36 & 4 \\
     & 2022-01-16 & 21:37:47 & 21:51:12 & 4 \\
     & 2022-02-18 & 19:07:30 & 19:21:08 & 4 \\
     & 2022-02-26 & 18:52:00 & 19:05:39 & 4 \\
     \rule{0pt}{1ex} \\
    EC & 2017-08-17 & 19:19:44 & 19:32:52 & 6 \\ %100/150s
    21077$-$4815 & 2022-06-07 & 23:12:24 & 23:28:19 & 6 \\ %75/100s
     & 2022-06-10 & 23:12:06 & 23:22:59 & 6 \\ %75/100s
     & 2022-06-25 & 22:39:51 & 22:49:21 & 6 \\ %75/100s
     & 2022-06-26 & 21:59:56 & 22:10:19 & 6 \\ %75/100s
     & 2022-06-27 & 22:18:01 & 22:28:25 & 6 \\ %75/100s
     & 2022-06-28 & 21:50:30 & 22:00:03 & 6 \\ %75/100s
     & 2022-06-30 & 22:23:12 & 22:37:34 & 6 \\ %30s
     \rule{0pt}{1ex} \\
    SB 705 & 2018-08-23 & 22:47:20 & 22:55:47 & 6 \\ %50/70s
     & 2022-10-30 & 18:37:45 & 18:51:16 & 6 \\ %60/90s
     & 2022-12-09 & 21:48:20 & 21:57:44 & 4 \\ %60/90s
     & 2022-12-29 & 21:01:38 & 21:07:50 & 4 \\ %60/90s
     & 2023-01-19 & 19:36:53 & 19:43:11 & 4 \\ %60/90s
     \rule{0pt}{1ex} \\
    Ton S 415 & 2020-11-02 & 22:25:21 & 22:31:44 & 4 \\
     & 2020-11-10 & 21:31:32 & 21:41:58 & 4  \\
     & 2022-08-16 & 03:16:40 & 03:53:33 & 21 \\
     & 2022-08-23 & 02:37:18 & 03:11:01 & 24 \\
     & 2022-08-29 & 02:14:36 & 02:30:05 & 12 \\
     & 2022-09-08 & 01:48:19 & 02:22:40 & 24 \\
     & 2022-09-11 & 01:19:49 & 01:53:42 & 23 \\
     \rule{0pt}{1ex} \\
     TYC & 2023-05-02 & 17:27:38 & 17:34:06 & 4 \\
     7242$-$541$-$1 & & & & \\
    \hline
    \end{tabular}
    \caption{Record of SALT observations for binary candidates.}
    \label{tab:obs_record}
\end{table}

SALT spectra for the identified candidates were used to search for evidence of periodic radial velocity shifts due to binary orbits. Spectra were obtained using SALT's Robert Stobie Spectrograph (RSS) with the PG2300 grating and a slit width of 1.0\,{\AA}, yielding a resolution $R=2250$ over a wavelength range of $3850-5150$\,{\AA} with a sampling of 0.34\,{\AA}/pix. Observing blocks consisted of two object frames and one CuAr arc frame each for grating angles of $30^{\circ}$ and $32.5^{\circ}$. Data reduction was performed using a modified version of the Titus Saures Rex pipeline\footnote{https://github.com/NaomiTitus} (Titus, priv. comm.). Cosmic ray removal was performed using the {\sc lacosmic} package \citep{vandokkum01}. The data reduction process is described in \citet{snowdonphd}.

\section{Methods}

\subsection{Light curve analysis}

The {\it TESS} Target Pixel Files (TPFs) for all candidates were obtained from the MAST portal. Light curves were extracted from the TPFs using {\sc lightkurve}. The default aperture masks defined by the {\it TESS} pipeline were used for the target stars. Aperture masks were also manually defined for any other bright objects in the TPF frame, as well as areas of background. Light curves were extracted using these masks to check for contamination of the target. Scattered light was removed using pixel level decorrelation \citep{deming15}. Finally, flux values more than $3\sigma$ from the mean were removed as outliers, and the light curves were normalised.

Lomb-Scargle periodograms were then computed from the processed light curves using {\sc lightkurve}'s to\_periodogram() function, which wraps {\sc astropy}'s LombScargle class \citep{vanderplas12, vanderplas15}. To ensure candidates showed strong evidence of close binarity, only signals which exceeded a $8\sigma$ threshold were considered potential periods. Period values were determined by fitting parabolas to the peaks in the periodogram. Each light curve was then phase-folded according to its period. 

Candidates were deemed to be contaminated if matching period signals could be seen more strongly in another bright object in the TPF frame. Candidates known to have bright objects lying within 1' were also considered potentially contaminated as the 21" {\it TESS} pixel size would make it difficult to resolve them from their neighbours.

\subsection{Spectrum radial velocity analysis}

Repeat observations of selected candidates were made using either the SALT Robert Stobie Spectrograph (RSS) or the SALT High Resolution Spectrograph (HRS) as for the survey observations described in \citet{jeffery20}. 

In most cases a single RSS observation consisted of 4 or 6 exposures at 2 grating angles. These were combined to form single normalised spectrum for each observation. 
The total wavelength coverage for each RSS observation is $3850-5100$\,{\AA} at a resolution of $3000$, which includes many He{\sc i}, He{\sc ii} and Balmer lines. 
For Ton S 415 repeat RSS observations were also made at a single grating angle to allow for an increased cadence; these are reported elsewhere \citep{snowdon23}. 

For EC 04110$-$1348, 12 repeat HRS observations were made. Each observation consisted of 2 exposures in the each of the blue and red arms. Data were reduced using standard {\sc iraf} routines \citep{snowdon22} and also using the SALT/HRS {\sc midas} pipeline \citep{kniazev16a,kniazev16b}. 
The pipeline reduced data were additionally rectified to the local continuum. 
Normally observed in pairs, we used all individual  HRS spectra observed with the blue arm.  

All radial velocity analysis was carried out using the {\sc starlink} package {\sc dipso} \citep{howarth04}. 
For each selected candidate a weighted mean spectrum was constructed from all available observations, where the weights were determined from the signal-to-noise ratio of individual observations. 
Radial velocities were measured relative to this mean spectrum using standard cross-correlation methods. 
Data were prepared for cross-correlation by resampling on a logarithmic wavelength scale, and subtracting the continuum.
The starting and ending 3\% of each spectrum were multiplied by a cosine bell to remove end effects, and the cross-correlation functions (CCF) computed from the result. 
The velocity and error is measured by fitting a Gaussian to the peak of the CCF. 

For RSS spectra, the entire wavelength range was used for the final CCF measurements. These were checked by examining CCFs covering restricted wavelength ranges including, for example, only He{\sc i}, only He{\sc ii}, or only weak lines. This allowed us to check for double components or components having different temperatures. No such evidence was found. 

For HRS spectra, subsections of the spectrum were used for CCF analysis, since the entire spectrum contained too many pixels for the analysis software to handle. Again, relative velocities from a few different wavelength ranges were compared to eliminate potential systematic effects. 
We note that the observations span over 6 months in time and hence the raw spectra include relative shifts of over 50\,kms$^{-1}$ due to the Earth's orbital acceleration. These were subtracted in the data reduction pipeline and were not detected in the CCF measurements. Since detector artefacts on SALT/HRS would follow the Earth velocity, the negative detection of any stellar acceleration cannot be due to instrumental artefacts.

\begin{table*}
\setstretch{1.25}
\caption{Summary of spectroscopic atmospheric parameters and results from the SED fit. Estimated uncertainties are stated for one sigma confidence. BPS CS 22956$-$0094 is a composite-colour binary with at least one K-type companion.}
\label{tab:specsed}
%\vspace{-12pt}
\begin{center}
\resizebox{\textwidth}{!}{
\begin{tabular}{l c c c c c c c c c}
\hline
   & \teff\  & $\log g$  & \logy   & $\log_{10} \theta$  &  $E$(44$-$55) & $\varpi$ & $R$ & $L$ & $M$\\
Star  & K  & $[\mathrm{cm}\,\mathrm{s}^{-1}]$  &     & [rad]  &   mag &  mas & ${\rm R_\odot}$ & ${\rm L_\odot}$ & ${\rm M_\odot}$\\

\hline
BPS CS 22956$-$0094 A  & $36800 \pm 1000$ & $6.05 \pm 0.15$ & $-0.03 \pm 0.05$ & $-11.046 \pm 0.0011$ & $0.023 \pm 0.005$ &  $1.82 \pm 0.12$ &  $0.110 \pm 0.008$ &  $20 \pm 4$ & $0.50^{+0.23}_{-0.16}$  \\
%BPS CS 22956$-$0094 B & $4450 \pm x$ &  $4.0$ (fix) & $-1$ (fix) &  $-x \pm 0.04$ & -- &  -- &  $x$ &  $x$ & $x$  \\
EC 04110$-$1348      & $47900 \pm 1200$ &  $5.66 \pm 0.11$ & $0.88 \pm 0.12$ &  $-10.875 \pm 0.004$ & $0.031 \pm 0.002$ &  $1.41 \pm 0.06$ &  $0.209 \pm 0.009$ &  $208^{+28}_{-25}$ & $0.74^{+0.22}_{-0.17}$  \\
EC 21077$-$4815      & $47100 \pm 1100$ &  $5.86 \pm 0.11$ & $1.17 \pm 0.18$ &  $-11.4240^{+0.0060}_{-0.0016}$ & $0.034 \pm 0.003$ &  $0.70 \pm 0.06$ &  $0.118\pm 0.010$ &  $63^{+13}_{-10}$ & $0.38^{+0.14}_{-0.10}$  \\ 
SB 705               & $47500 \pm 1100$ &  $5.77 \pm 0.11$ & $1.17 \pm 0.12$ &  $-10.979 \pm 0.004$ & $0.010 \pm 0.002$ &  $1.13 \pm 0.05$ &  $0.206 \pm 0.010$ &  $196^{+26}_{-23}$ & $0.91^{+0.27}_{-0.21}$  \\
Ton S 415$^\dagger$  & $43330 \pm1 000$ &  $5.89 \pm 0.10$ & $-0.62\pm0.10$  &  $-11.038^{+0.006}_{-0.005}$ & $0.012 \pm 0.002$ &  $1.89 \pm 0.04$ &  $0.1074^{+0.0025}_{-0.0024}$ &  $37 \pm 4$ & $0.33^{+0.09}_{-0.07}$  \\
TYC 7242$-$541$-$1   & $51600 \pm 1400$ &  $5.83 \pm 0.11$ & $1.45^{+0.22}_{-0.12}$ &  $-10.834 \pm 0.005$ & $0.049 \pm 0.003$ &  $1.65 \pm 0.07$ &  $0.196 \pm 0.009$ &  $250 \pm 40$ & $0.96^{+0.27}_{-0.22}$  \\
\hline
\end{tabular}
}
\end{center}
%\vspace{-12pt}
%\begin{tabular}{P{175mm}}
    \parbox{175mm}{$^\dagger$ from \citet{snowdon23}. 
    }\\
%\end{tabular}
\end{table*}

\subsection{Spectral analysis}

For stars deemed to be {\it bona fide} variables we have measured their overall properties by analysis of their RSS spectra and of their spectral energy distributions (SEDs). 

\begin{figure*}
    \centering
    \includegraphics[width=\textwidth]{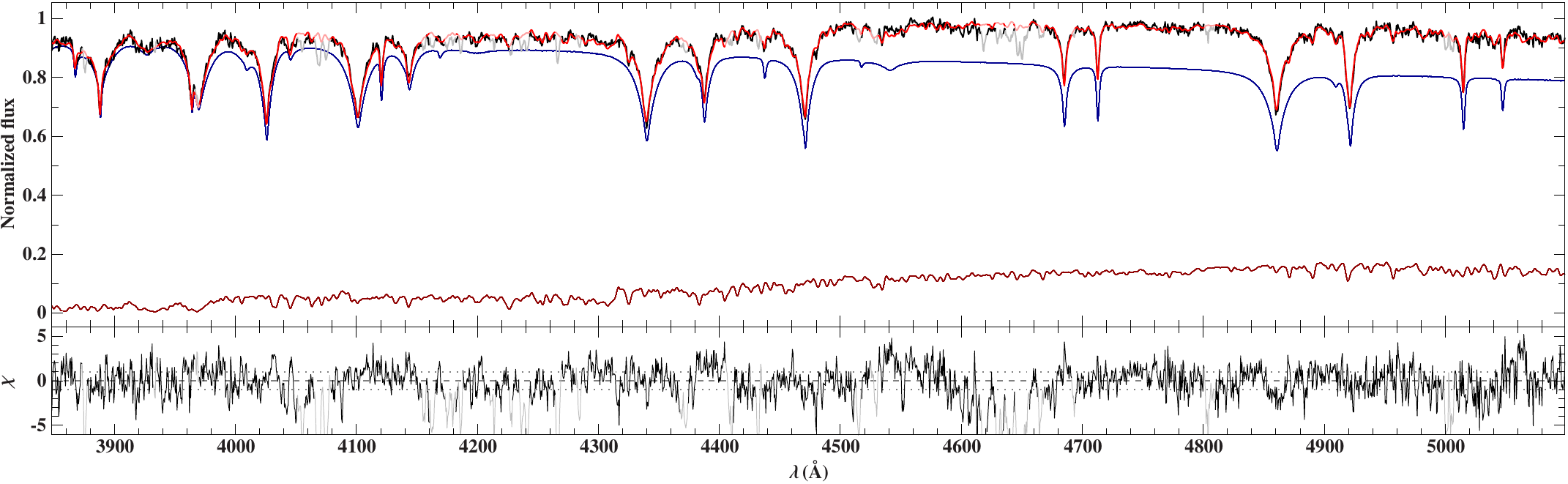}
    \caption{Best fit (red) to an RSS spectrum of BPS CS 22956-0094 (black). The model consists of a hot subdwarf (blue) and a K-type component (dark red). The bottom panel shows the uncertainty-weighted residuals $\chi$ between model and observation. 
    }
    \label{fig:bps_spec}
\end{figure*}

We used the model grid of \cite{dorsch21} to obtain atmospheric parameters \teff, \logg\ and \logy\ from $\chi^2$ minimisation\footnote{These fits were performed in the \textsc{isis} framework \citep{Houck2000} by the method of \cite{Irrgang2014}.} to these spectra. This large model grid was calculated using \textsc{Synspec}, based on \textsc{Tlusty} model atmospheres in non-local thermodynamic equilibrium \citep{Hubeny2017a}. Since 2021, it has been extended to cover the full He-sdO parameter range. 
In case multiple spectra were available, all spectra were fit individually and the parameters reported in Table \ref{tab:specsed} to average values. 
Our RSS spectra show no evidence of contributions from secondary components, with the exception of BPS CS 22956-0094. 
This system shows strong contribution by both an intermediately helium-rich sdOB star and a K-type star, as shown in Fig.\ \ref{fig:bps_spec}. 
We therefore modeled this system using two components -- the sdOB was represented by the \textsc{Tlusty}/\textsc{Synspec} model grid, while the cool component was modeled using \textsc{Atlas12}/\textsc{Synthe} \citep{Kurucz1993, Kurucz1996}, the same method as used by \citet{dorsch21}. 
Since our main interest lies in the sdOB component, we only state the parameters of this component in Table \ref{tab:specsed}. A more detailed analysis of this system should be performed in a future study.  

These parameters can be combined with the \textit{Gaia} DR3 \citep{gaia2023} parallax $\varpi$ and an angular diameter $\Theta$ from photometric flux measurements to derive radii $R = \Theta/(2\varpi)$ and luminosities $L = 4\pi R^2 \sigma_\mathrm{SB} T^4_\mathrm{eff}$, where $\sigma_\mathrm{SB}$ is the Stefan–Boltzmann constant. 
To this end we performed SED fits, using \textit{Gaia} DR3 parallaxes with  zero-point corrections applied according to \cite{Lindegren2021} and  inflated uncertainties following \cite{El-Badry2021}. 
Stellar masses may be estimated via $M = g R^2/G$ but remain poorly constrained due to uncertainty in the surface gravity $g$. 
While the stellar masses are not well determined, the derived radii and luminosities are quite precise. 

Table \ref{tab:specsed} summarises the results of our spectral and SED fits. 

\section{Results}

\subsection{Rejected candidates}

Upon inspection, apparent period signals in the light curves of 6 candidates were found to be likely caused by contamination. EC 20081$-$3205 was found to be contaminated by the nearby bright star UCAC4 291$-$219126, a W UMa eclipsing contact binary \citep{drake17}. [CW83] 0832$-$01 was similarly contaminated by the nearby star TIC 121318598. The default aperture masks for BPS CS 22885$-$0043 and EC 20577$-$5504 were found to be in the wrong place on their TPF frames. When moved to cover the actual targets, the apparent variability disappeared from the periodograms. UCAC4 204$-$139111 was rejected due to its very low CROWDSAP value of 0.2265. EC 22536$-$5304 was found to have a bright star within $\sim40"$ which could not be fully resolved from the target. EC 22536$-$5304 is also known to be a long-period binary with a metal-poor subdwarf F-type companion \citep{dorsch21}.

In addition to the candidates excluded due to contamination, GALEX J084528.7$-$12140 was rejected due to poor data quality in the sector 34 TPF, as well as the discovery of multiple bright objects lying within 1' which would not be resolvable by {\it TESS} pixels. PG 2158$+$082 was similarly rejected due to the presence of unresolvable nearby stars. HD 124448 (aka V821 Cen) was excluded as it is known to be a pulsating extreme helium star \citep{jeffery20d}. Finally, EC 04399$-$1826 was rejected due to the very small amplitude and poor signal-noise in its folded light curve.

\subsection{Identified variables}

Of the initial 16 candidates identified from {\it TESS} data, only Ton S 415 was confirmed to be a close binary system. The star showed two peaks in its periodogram at $84.6$ and $42.3\,{\rm min}$, as well as clear periodic radial velocity variations in its SALT spectra with a large semi-amplitude of $175.5\pm1.0\,{\rm kms}^{-1}$. Due to the 2:1 frequency ratio of the periodogram signals, combined with the lack of evidence of a cool companion in the SED, it was concluded that Ton S 415 is in a close binary with an unseen white dwarf companion, which causes ellipsoidal deformation of the subdwarf primary. The full analysis of Ton S 415 is described by \citet{snowdon23}.

\begin{figure*}
    \centering
    \includegraphics[width=\textwidth]{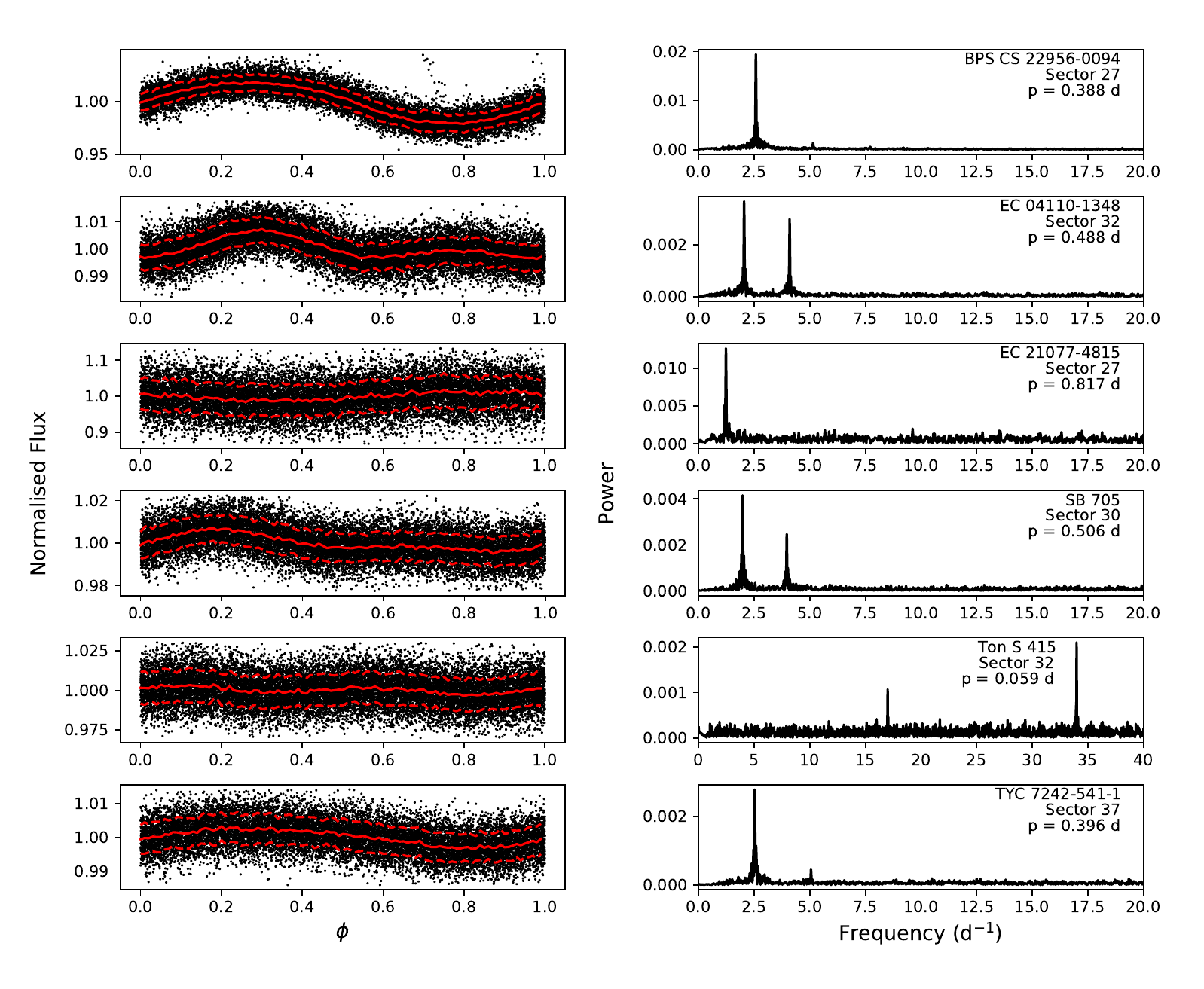}
    \caption{(Left) example {\it TESS} light curves of identified SALT survey variable stars, normalised and phased to the strongest period signal. The mean and standard deviation of the flux across 100 bins is overlaid in red. (Right) periodograms computed from the light curves, the {\it TESS} sector numbers and periods ($p$) corresponding to the highest-power signals are indicated.
    }
    \label{fig:lcpgs}
\end{figure*}

Sample light curves and periodograms of the remaining candidates are shown in Figure \ref{fig:lcpgs}. 

\begin{figure*}
     \centering
     \begin{subfigure}[b]{0.95\columnwidth}
         \centering
         \includegraphics[width=\textwidth]{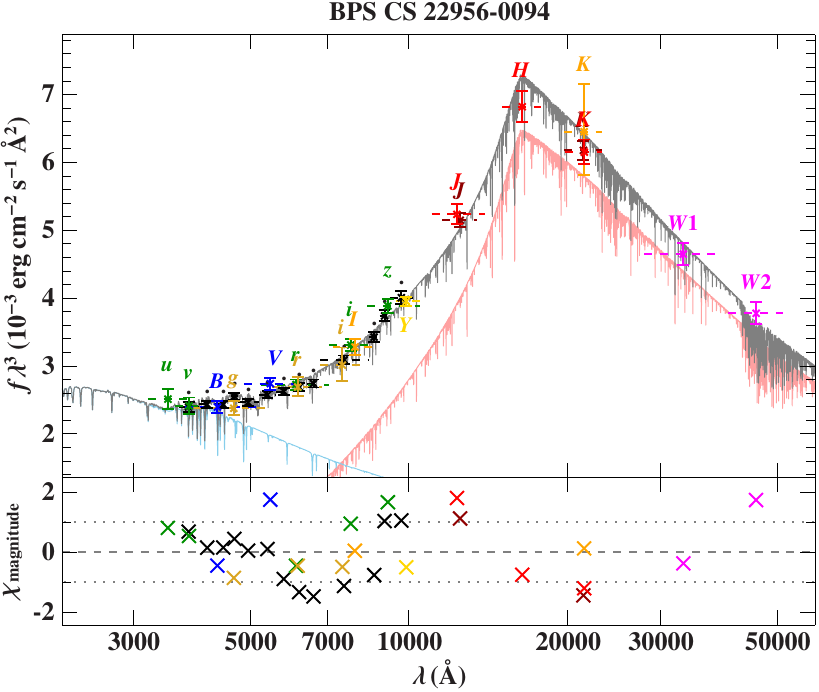}
     \end{subfigure}
     \hfill
     \begin{subfigure}[b]{0.95\columnwidth}
         \centering

         \includegraphics[width=\textwidth]{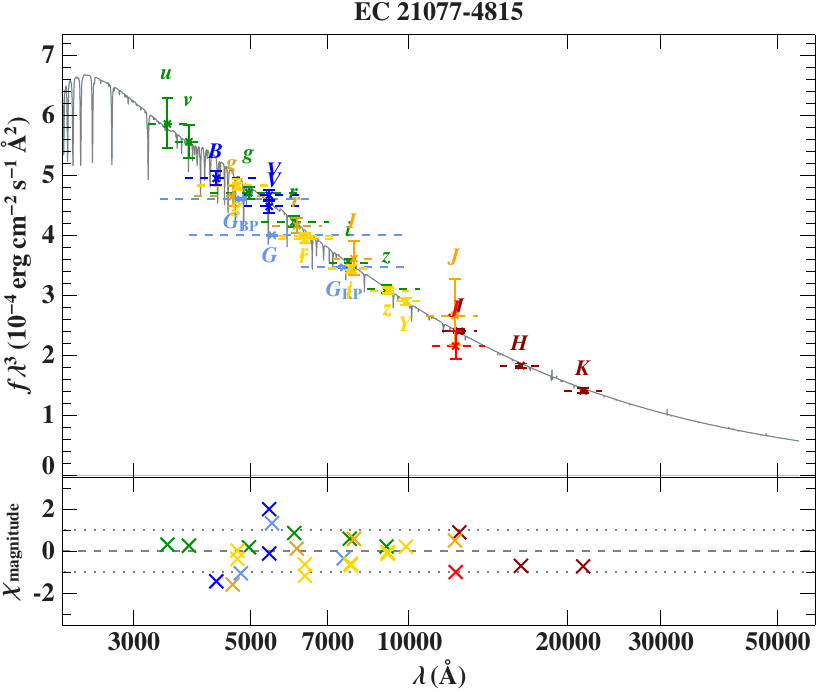}
     \end{subfigure}
        \caption{Spectral energy distribution for BPS CS 22956$-$0094 (left), where the infrared excess indicative of a cool companion (red) can be clearly seen. The hot subdwarf contribution is shown in blue. The SED for EC 210774$-$4815 (right) is shown for comparison and is typical of the other stars in the sample.}
        \label{fig:seds}
\end{figure*}

BPS CS 22956$-$0094 shows clear variability in all its {\it TESS} sectors at a period of 0.388\,d, with the highest signal-noise light curve in the sample. Of the 6 identified variables, this is the only star to show any evidence of a red or infrared excess in its SED (Fig. \ref{fig:seds}). This star might be a triple containing the He-sdOB and two K-type companions, or a blend. This is because the radius ($1.1\,{\rm R_{\odot}}$) and luminosity ($0.42\,{\rm L_{\odot}}$) obtained from the SED fit for a single K-type companion are too large for a MS K-type and too small for a base of the RGB star. There is no track that matches a single K-type companion. If there are two K-types, the fit to evolutionary tracks is better, but still not good; the companions are still too large. None of this affects the fit for the He-sdOB.
No strong evidence of radial velocity shifts could be found from the 10 available RSS spectra, making it unlikely that BPS CS 22956$-$0094 and its companion are in a close binary.

EC 21077$-$4815 shows variability with a period of 0.815\,d in both available sectors, though with a poorer signal-noise and CROWDSAP compared to BPS CS 22956$-$0094. However, the star shows neither radial velocity variations nor evidence for a companion in the SED.

SB 705 and EC 04110$-$1348 have very similarly-shaped, double-peaked, asymmetric light curves. Both stars also have two signals in their periodograms at periods of $\sim0.25\,{\rm d}$ and $\sim0.5\,{\rm d}$. Neither star shows any significant radial velocity variations. 

TYC 7242$-$541$-$1 shows similar variability to BPS CS 22956$-$0094, with a period of $0.396\,{\rm d}$ and a small harmonic at around $0.2\,{\rm d}$. However, it does not show any evidence of a cool companion in its SED. Insufficient SALT spectra were available to determine radial velocity variability.

\begin{figure}
    \centering
    \includegraphics[width=\columnwidth]{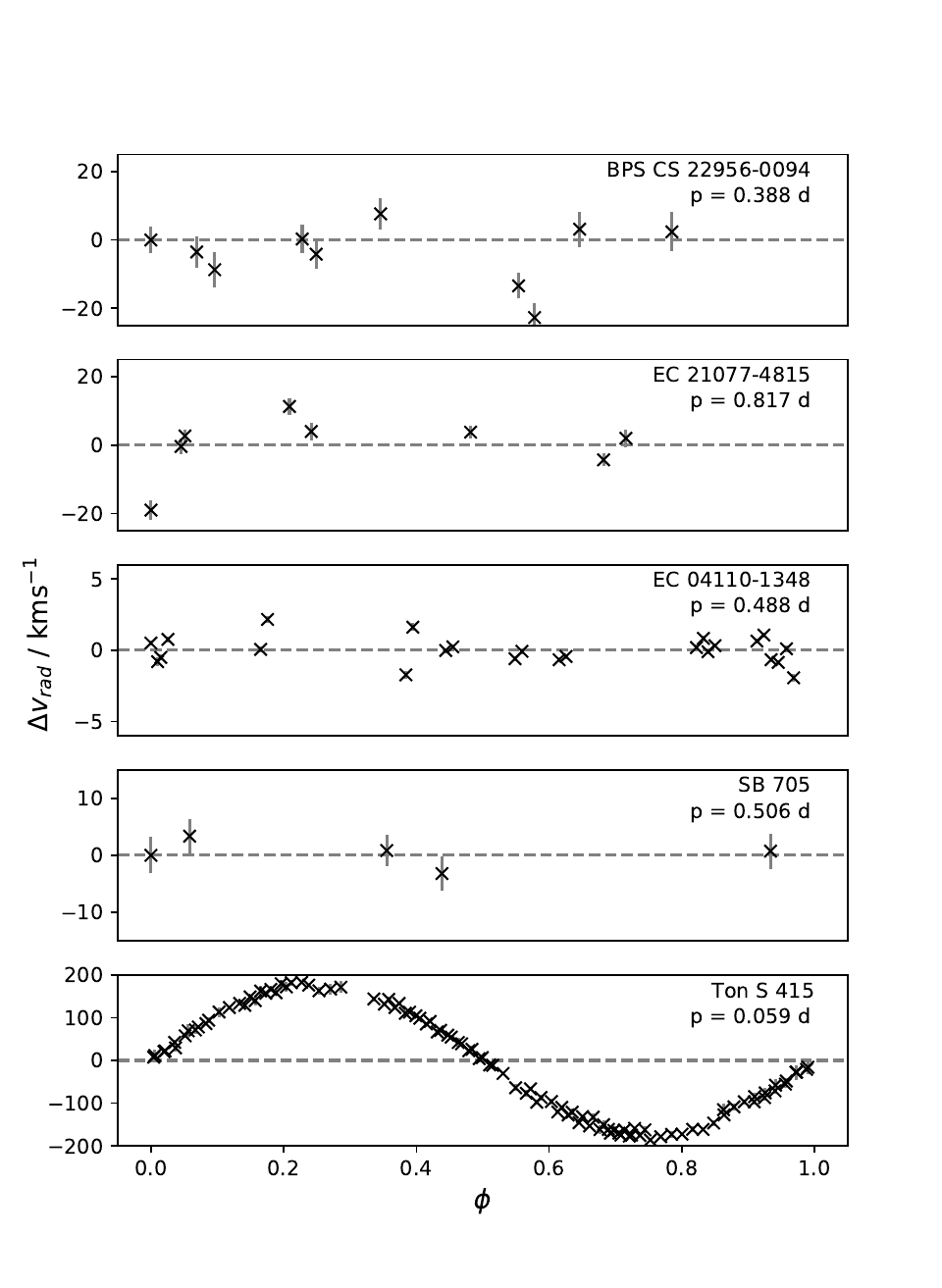}
    \caption{Radial velocity measurements from cross-correlation for the SALT candidate binaries. Each star's measurements have been phase-folded to their periods in Table \ref{tab:tess_objects}. TYC 7242$-$541$-$1 is excluded due to a lack of available spectra. Apart from Ton S 415, no candidates show significant radial velocity variations corresponding to their photometric periods. The formal error in  each RV measurement is shown in grey. The errors for the HRS measurements of EC 04110$-$1348 are especially small, as are those for Ton S 415 relative to its amplitude.}
    \label{fig:vrads}
\end{figure}

With the exceptions of Ton S 415 (iHe-sdOB) and BPS CS 22956$-$0094 (iHe-sdB), all of the identified variables were He-rich sdO stars. This may reflect an observational bias, as He-sdOs form a significant majority of the hydrogen-deficient SALT sample, as seen in Fig.\ref{fig:saltsurveyclasses}.

\section{Discussion}

Out of 153 hydrogen-deficient SALT survey catalogue objects with available {\it TESS} data only a single short-period binary system, Ton S 415, could be conclusively identified. The fraction of confirmed hydrogen-deficient close binaries in the SALT survey is therefore $\sim0.65\pm0.65\%$. This is similar to the fraction given by \citet{geier22}, who list 5 known solved close binaries out of a sample of approximately 600 for a fraction of $\sim0.83\pm0.38\%$. These binaries comprise:

\begin{itemize}
    \item PG 1544$+$488, a double-lined He-sdB + He-sdB binary \citep{ahmad04}
    \item Hen 2$-$248, a double-lined iHe-sdOB + sdOB binary \citep{reindl20}
    \item CPD $-20^{\circ}1123$, a single-lined iHe-sdB with an unseen companion \citep{naslim12}
    \item OW J0741$-$2948, a highly compact iHe-sdOB + WD binary \citep{kupfer17}
    \item ZTF J2130$+$4420, a highly compact iHe-sdOB + WD binary \citep{kupfer20}
\end{itemize}

Of these, Ton S 415 most closely resembles OW J0741$-$2948 and especially ZTF J2130$+$4420, though Ton S 415 does not appear to have synchronised its rotation to its orbit, and does not show evidence of accretion onto the companion. These three stars have periods $\leq 90\,{\rm min}$ and intermediate He abundances. \citet{kupfer20} proposed that these objects form from ignition of non-degenerate core helium burning in relatively massive progenitor stars, producing hot subdwarfs with sub-canonical masses and moderate He enrichments in compact binaries. 

The apparent presence of a K-type companion in the SED of BPS CS 22956$-$0094, combined with the lack of significant radial velocity variations, suggests that the system is likely a detached binary. Determination of the true orbital period would require additional spectra to be obtained over a longer timeframe. Hot subdwarf-main sequence binaries with periods as long as several hundred days can form via stable Roche lobe overflow in progenitor systems with a mass ratio of approximately $1.2-1.5$ \citep{podsiadlowski08, vos18}. 

Compared to known classes of sdB pulsators, the period of $0.388\,{\rm d}\simeq33\,500\,{\rm s}$ is much longer than the $80-580\,{\rm s}$ periods seen in V361 Hya stars \citep{romero21}. With an effective temperature of $36\,900\,{\rm K}$, the star is also significantly hotter than the population of longer-period g-mode pulsators, which have $T_{\rm eff}<30\,000\,{\rm K}$ \citep{reed04}. Since {\it TESS} is more sensitive to redder wavelengths than SALT, it may be that the cool companion is the actual source of variability. Pulsations are unlikely, as K-dwarfs do not show large amplitude global oscillations. Variability due to rotation of surface starspots would require a high rotational velocity of approximately $130\,{\rm kms}^{-1}$ for a solar-radius star with a period of 0.388 days. Alternatively, the variations could be caused by orbital motion between the K-dwarf and an unseen third component to the system. High-resolution spectroscopy of the companion will be needed to investigate these possible scenarios.

EC 21077$-$4815 can be ruled out as a binary star due to the lack of radial velocity variations in the spectral lines or a companion in the SED. All known sdO pulsators have p-mode pulsations with periods on the order of minutes, significantly shorter than the $0.817\,{\rm d}$ period seen in EC 21077$-$4815 \citep{baran23}. The comparatively low signal-noise of the light curve is likely a result of EC 21077$-$4815 being the faintest star out of the identified candidates ($m_{\rm G}=15.2$). The star also has a relatively poor CROWDSAP value of 0.79, though inspection of the TPF did not suggest contamination from other bright objects in the frame. A possible explanation could be the presence of an unseen low-mass companion. Brown dwarfs and hot subdwarfs have similar radii, so their relative intrinsic brightnesses go according to $(T_{\rm BD}/T_{\rm sd})^4$. The flux ratio would not exceed $\sim1\%$, which would not be detectable in the SED. A $10\%$ flux variation in a hypothetical BD companion would therefore be $<0.1\%$ of the subdwarf flux, which is much smaller than the observed variation. The H and K magnitudes in the SED preclude the presence of companions more massive than about $0.2\,{\rm M_{\odot}}$. A MS companion below this mass limit should produce a RV semi-amplitude of $10\leq K \leq100\,{\rm kms}^{-1}$ (per Fig. 18 in \citet{schaffenroth23}). Assuming a companion with $M=0.2\,{\rm M_{\odot}}$, $R=0.2\,{\rm R_{\odot}}$, EC 21077$-$4815 would require a nearly pole-on inclination angle of $\geq 80^{\circ}$ in order to match both the observed photometric variability and lack of radial velocity variations.

EC 04110$-$1348 and SB 705 show remarkable similarities in the shapes of their light curves and the corresponding periods. The stars are also very similar spectroscopically, with Drilling classes of sdOC7.5VIIHe:39 and sdOC7.5VIIHe:40 respectively \citep{jeffery20}. The 2:1 frequency ratio of the peaks in the periodograms could suggest ellipsoidal deformation by an unseen companion, as seen in Ton S 415. However, unlike Ton S 415, neither star displays significant radial velocity variations. The asymmetric, double-peaked shape of the light curve also presents a problem. Doppler beaming could cause a similar asymmetry, but would require large projected orbital velocities which are not seen in the radial velocity measurements. One possible explanation could be that the variations are caused by chemical bright spots on the stellar surfaces. Spot-induced variability was found in globular cluster EHB stars by \citet{momany21}, though these objects are significantly cooler than EC 04110$-$1348 and SB 705 and also have longer periods of $>2\,{\rm d}$. For the variation to be caused by spot rotations, both stars would need a rotational velocity of approximately $20\,{\rm kms}^{-1}$ (for assumed stellar radii of $0.2\,{\rm R_{\odot}}$ and periods of $0.5\,{\rm d}$). The resulting line broadening would be $<1\,${\AA} in the SALT wavelength range and thus not detectable using the resolution of RSS. Higher-resolution spectral data will be needed to confirm whether or not the stars' rotational velocities match the observed periods. EC 21077$-$4815 (sdOC7.5VII:He39) is also a close spectroscopic match to both EC 04110$-$1348 and SB 705, suggesting spots as a potential explanation for its variability as well. However, its longer period of $0.815\,{\rm d}$ would result in even weaker rotational broadening. Coupled with EC 21077$-$4815 being $\sim2$ magnitudes fainter than the other stars, it could be difficult to spectroscopically confirm the presence of surface bright spots.

TYC 7242$-$541$-$1 cannot be conclusively confirmed or rejected as a close binary candidate until its radial velocity variability is determined, which will require further observations by SALT. Although it has a similar periods and shape of light curve to BPS CS 22956$-$0094, it is not spectroscopically similar and shows no evidence of a cool companion in its SED. As such, the potential non-orbital explanations for the variability of BPS CS 22956$-$0094 such as pulsation or the rotation of a spotted companion are not necessarily valid for TYC 7242$-$541$-$1.

\section{Conclusions}

Analyses of the available {\it TESS} light curves for the SALT catalogue of hydrogen-deficient stars identified 16 candidate variables, of which 10 were later rejected due to contamination or poor data quality. Radial velocity and SED analyses of the remaining candidates identified one confirmed close binary, Ton S 415. This places the known fraction of close binaries in the SALT catalogue at 1 in 153 objects, or $\sim0.65\%$. This can be further divided into fractions $1/29\simeq3.45\%$ for intermediate helium objects and $0/124 = 0\%$ for extreme helium objects. This is in line with the low binary fractions predicted by the formation channels for hydrogen-deficient stars, but may be an underestimate due to observational biases and the limitations of {\it TESS} data.

The source of the variability in the remaining 5 candidates remains unknown. BPS CS 22956$-$0094 may be in a long-period binary, indicated by the presence of a K-type companion in its SED but lack of accompanying periodic radial velocity variations. EC 21077$-$4815 does not show any evidence of a binary companion, and its variability does not match expectations for a pulsating sdO. As a faint star with photometry available from only 2 {\it TESS} sectors, the available data is not sufficient to conclusively determine the cause of variability. EC 04110$-$1348 and SB 705 display remarkably similar variations while also showing no evidence of having binary companions. Chemical bright spots on the stellar surfaces could possibly account for the variations, but the resolution of the available RSS spectra is not sufficient to determine if this is the case. TYC 7242$-$541$-$1 requires additional observations from RSS and/or HRS to assess radial velocity variability before it can be confirmed or rejected as a close binary candidate.

\section*{Acknowledgements}

The authors wish to thank Dr. Naomi Titus for providing the basis for the data reduction pipeline and Mr. Asish Philip Monai for assisting with the record of SALT observations.

Some of the observations reported in this paper were obtained with the Southern African Large Telescope (SALT) under program 2021-1-MLT-005 (PI Jeffery). 
This paper includes data collected by the {\it TESS} mission. Funding for the {\it TESS} mission is provided by the NASA's Science Mission Directorate. 
EJS is supported by the United Kingdom Science and Technology Facilities Council (STFC) via UK Research and Innovations (UKRI) doctoral training grant ST/R504609/1. 
CSJ is supported by the STFC via UKRI research grant ST/V000438/1. 
The Armagh Observatory and Planetarium (AOP) is funded by direct grant from the Northern Ireland Department for Communities. This funding also provides for AOP membership of the United Kingdom SALT consortium (UKSC). 
SS acknowledges financial support from AOP in the form of an internship organised by the International Association for the Exchange of Students for Technical Expertise (IAESTE).
For the purpose of open access, the authors have applied a Creative Commons Attribution (CC BY) license to any Author Accepted Manuscript version arising.

%%%%%%%%%%%%%%%%%%%%%%%%%%%%%%%%%%%%%%%%%%%%%%%%%%
\section*{Data Availability}

A copy of the SALT data may be made on reasonable application to the authors. The SALT data will be available from the SALT data archive (https://ssda.saao.ac.za) after the proprietary period ends in April 2025. {\it TESS} photometric data are available through the MAST portal (https://mast.stsci.edu).


\bibliographystyle{mnras}
\bibliography{saltbinaries}

%%%%%%%%%%%%%%%%%%%%%%%%%%%%%%%%%%%%%%%%%%%%%%%%%%

%%%%%%%%%%%%%%%%% APPENDICES %%%%%%%%%%%%%%%%%%%%%

\appendix
\section{SALT stars inspected with {\it TESS}}\label{app:tess}

\begin{table*}
    \setlength{\tabcolsep}{3pt}
    \begin{tabular}{lllc|lllc}
    \hline
    \multicolumn{8}{c}{Other Stars}\\
    \hline
    {\it Gaia} DR3 & TIC & Name & Sectors & {\it Gaia} DR3 & TIC & Name & Sectors\\
	6570203444846881536 & 278705842 & BPS CS 22875$-$0002 & 2 & 2909401745277447808 & 153165824 & GLX J055804.5$-$292708 & 3\\
	6825916554879185664 & 409119442 & BPS CS 22892$-$0051 & 2 & 2899570977454318336 & 37602756 & GLX J061237.5$-$271254 & 2\\
	6394226540100907904 & 403317759 & BPS CS 22938$-$0044 & 4 & 5286825950756440832 & 348992108 & GLX J070738.9$-$622241 & 33\\
	6393692417967690240 & 237338096 & BPS CS 22938$-$0073 & 3 & 5491580506476906752 & 294394152 & GLX J071549.6$-$540755 & 8\\
	6405416373015926656 & 389078407 & BPS CS 22956$-$0090 & 5 & 3068507812327598208 & 25948892 & GLX J075807.5$-$043203 & 3\\
	2414888694102153088 & 439397980 & BPS CS 29517$-$0049 & 3 & 5730001395279470592 & 347412256 & GLX J085158.8$-$171238 & 4\\
	6804485252188569728 & 99499703 & BPS CS 30319$-$0062 & 2 & 5433906762213163392 & 45362220 & GLX J095256.6$-$371940 & 5\\
	4921084329876536448 & 281741309 & EC 00468$-$5440 & 1 & 3642867535345067392 & 287959455 & GLX J142549.8$-$043231 & 1\\
	4927860005859630080 & 616164061 & EC 01086$-$5138 & 1 & 5823037151488582272 & 342238727 & GLX J160011.8$-$643330 & 3\\
	4911021775553809664 & 206590489 & EC 01196$-$5548 & 3 & 5804428127642370944 & 301454053 & GLX J170506.0$-$715609 & 3\\
	4646812876610294144 & 259963735 & EC 02523$-$6934 & 6 & 6629436197824489984 & 303537262 & GLX J180818.6$-$650853 & 2\\
	5099182265566440320 & 92358656 & EC 03263$-$2109 & 1 & 6629448391230142464 & 365432876 & GLX J181356.9$-$652747 & 3\\
	4667419442501873408 & 32068755 & EC 03505$-$6929 & 24 & 6705178908161092352 & 91316983 & GLX J183231.7$-$474435 & 2\\
	4844000803764527360 & 369377398 & EC 04013$-$4017 & 3 & 4046679845165833472 & 380274928 & GLX J183716.7$-$312514 & 1\\
	4891033684953481600 & 178873605 & EC 04271$-$2909 & 3 & 6651115985843036800 & 120122222 & GLX J183845.6$-$540935 & 2\\
	4789068103329323648 & 153835390 & EC 04281$-$4738 & 3 & 6716376055399759360 & 253656464 & GLX J184559.8$-$413827 & 1\\
	4677213647268641152 & 471013513 & EC 04369$-$6203 & 29 & 6711368192251441408 & 61029108 & GLX J190555.7$-$443838 & 3\\
	4877454200954089088 & 170869314 & EC 04405$-$3211 & 2 & 6711206632764706688 & 61379646 & GLX J191049.5$-$441713 & 3\\
	4818530857425524736 & 77477081 & EC 04517$-$3706 & 3 & 6438915331219654400 & 464653914 & GLX J191223.5$-$624632 & 3\\
	4761967615605547264 & 685125879 & EC 04534$-$5931 & 21 & 6713132053718884480 & 61621625 & GLX J191504.3$-$423502 & 3\\
	2906636679691633920 & 31174073 & EC 05242$-$2900 & 3 & 6746304422609405696 & 113118441 & GLX J193046.0$-$305000 & 3\\
	5494762428704591488 & 350843318 & EC 05593$-$5901 & 21 & 6688196710548976000 & 425799931 & GLX J193740.3$-$430356 & 3\\
	5685954371058159616 & 293603085 & EC 09557$-$1551 & 2 & 6880570941438210176 & 14348341 & GLX J201318.8$-$120119 & 1\\
	5457435108454471424 & 188174892 & EC 10475$-$2703 & 3 & 6906152621346929280 & 157440904 & GLX J202506.0$-$080419 & 1\\
	5457394701402238336 & 188207457 & EC 10479$-$2714 & 2 & 6914530384555510656 & 248105981 & GLX J203900.4$-$044241 & 1\\
	3544861947929180544 & 219249015 & EC 11236$-$1945 & 2 & 6588673900161571584 & 152286180 & GLX J215759.6$-$333506 & 3\\
	3471908183194405120 & 60709794 & EC 12349$-$2824 & 2 & 6358022233538484480 & 290401825 & GLX J220151.3$-$755202 & 2\\
	3495325684921885184 & 9335707 & EC 12420$-$2732 & 2 & 6516769619278068864 & 161153327 & GLX J223521.9$-$502119 & 3\\
	6292958495525136768 & 32145732 & EC 13290$-$1933 & 2 & 6543233004436306432 & 160078278 & GLX J230525.4$-$404634 & 3\\
	6642621850497112192 & 166224562 & EC 19371$-$5253 & 3 & 2336589275632410496 & 12378718 & HE 0001$-$2443 & 3\\
	6694364283585300736 & 355051924 & EC 20111$-$3724 & 2 & 2316462680926136960 & 63695388 & HE 0016$-$3212 & 3\\
	6423111535197509120 & 374408958 & EC 20111$-$6902 & 3 & 2455641680268793472 & 439404889 & HE 0111$-$1526 & 1\\
	6696646423047419264 & 34520844 & EC 20184$-$3435 & 2 & 4968603165141249792 & 152318833 & HE 0155$-$3710 & 2\\
	6667819869569857920 & 166578649 & EC 20187$-$4939 & 1 & 5058959140928424448 & 209452383 & HE 0301$-$3039 & 1\\
	6679173843252572672 & 129580709 & EC 20193$-$4339 & 2 & 4779256473879923072 & 198077858 & HE 0414$-$5429 & 6\\
	6429967711751930368 & 466310972 & EC 20221$-$6249 & 2 & 3485687950109280896 & 14717912 & HE 1136$-$2504 & 2\\
	6455294377981864320 & 220347928 & EC 20262$-$6000 & 2 & 3574711931980454528 & 152823245 & HE 1203$-$1048 & 3\\
	6474852834412549632 & 100475285 & EC 20306$-$5127 & 2 & 6318694764197053952 & 79286767 & HE 1511$-$1103 & 1\\
	6798823042186363648 & 212352349 & EC 20384$-$2816 & 2 & 4779307463731785088 & 396720998 & LB 1721 & 6\\
	6376187436940164480 & 372172495 & EC 20450$-$6947 & 3 & 4777646032942511872 & 259392878 & LB 1741 & 7\\
	6795812407549550080 & 471013483 & EC 20451$-$2828 & 1 & 4782839331302068224 & 220459208 & LB 1766 & 12\\
	6470065629505057152 & 354134858 & EC 20481$-$5518 & 2 & 4916399452565698688 & 229144939 & LB 3229 & 5\\
	6476519109924854528 & 166657780 & EC 20535$-$5251 & 2 & 5306775455393745024 & 386653873 & LSS 1274 & 5\\
	6457665096848737280 & 219304704 & EC 20577$-$5641 & 2 & 5985819027688073856 & 189834632 & LSS 3378 & 1\\
	6801342332560627200 & 29821374 & EC 21013$-$2723 & 3 & 2626098280727562496 & 439818822 & PB 7124 & 1\\
	6375472376424837248 & 419571221 & EC 21125$-$7013 & 3 & 2514184283536110720 & 270380474 & PG 0208$+$016 & 6\\
	6562925846101484544 & 139754135 & EC 21306$-$4911 & 4 & 6052403489630720 & 318804768 & PG 0240$+$046 & 3\\
	6586406672826522240 & 197573462 & EC 21416$-$3645 & 2 & 580372234855320064 & 270562073 & PG 0902$+$057 & 7\\
	6385665467688866304 & 325175734 & EC 22332$-$6837 & 4 & 5758956484239854336 & 170888068 & PG 0914$-$037 & 1\\
	6384973977954481792 & 469978554 & EC 22450$-$6827 & 5 & 3766459205017226240 & 332701732 & PG 0958$-$119 & 2\\
	2383525983912887168 & 204327923 & EC 23062$-$2348 & 2 & 3799987433421443712 & 380154683 & PG 1127$+$019 & 3\\
	6534198699642972160 & 183373104 & EC 23477$-$4130 & 3 & 3584539233766055296 & 147696519 & PG 1220$-$056 & 1\\
	6494992795056667648 & 355638102 & EC 23507$-$5733 & 5 & 3708604861669017728 & 399087019 & PG 1230$+$067 & 2\\
	3253550426661848960 & 9268360 & GLX J041110.1$-$004848 & 2 & 3709293602624709760 & 390687074 & PG 1248$+$066 & 2\\
	3255780171819962496 & 454147106 & GLX J042034.8$+$012041 & 2 & 3717383603022961792 & 393670835 & PG 1318$+$062 & 1\\
	2975366979747110016 & 398801916 & GLX J051348.2$-$194417 & 2 & 3717793239823667200 & 365070057 & PG 1325$+$054 & 1\\
    \hline
    \end{tabular}
    \caption{Summary of stars from the SALT survey which did not show variability in their {\it TESS} periodograms above an $8\sigma$ threshold.}
    \label{tab:app2}
\end{table*}

\begin{table*}
    \setlength{\tabcolsep}{3pt}
    \begin{tabular}{lllc|lllc}
    \hline
    \multicolumn{8}{c}{Other Stars}\\
    \hline
    {\it Gaia} DR3 & TIC & Name & Sectors & {\it Gaia} DR3 & TIC & Name & Sectors\\
	4421118770476595200 & 371354672 & PG 1528$+$029 & 1 & 6112506922060551168 & 166322623 & SALT J134648.6$-$402538 & 2\\
	4400811306108496000 & 46357457 & PG 1537$-$046 & 1 & 6095417723923269760 & 275921038 & SALT J135237.7$-$463925 & 4\\
	2691204212499446272 & 376401628 & PG 2116$+$008 & 1 & 5854075647522443904 & 331665132 & SALT J141007.4$-$623025 & 2\\
	2678666962443647232 & 439816832 & PG 2213$-$006 & 2 & 5815599887346246144 & 363740394 & SALT J163947.4$-$663152 & 1\\
	2652331593079280000 & 310937782 & PG 2303$+$013 & 2 & 5811865018141018752 & 301899016 & SALT J171737.4$-$670154 & 1\\
	2644373607090404480 & 398584811 & PG 2333$-$002 & 2 & 4233662070991147520 & 439763808 & SALT J195450.9$-$030732 & 1\\
	6817299093841252864 & 25300088 & PHL 178 & 3 & 2629387431107687168 & 2024377279 & SALT J222027.3$-$020134 & 2\\
	2638471188154451584 & 2052695518 & PHL 417 & 2 & 6549441534281786624 & 175295953 & SALT J230901.4$-$392954 & 1\\
	2437994896599568000 & 49530055 & PHL 443 & 1 & 2330920159319600000 & 254287117 & TON S 103 & 3\\
	2437293614339507968 & 49711130 & PHL 540 & 4 & 2335246080444203520 & 12423805 & TON S 144 & 3\\
	3087687074686316416 & 266480316 & SALT J075448.2$+$014612 & 2 & 3206674676854713344 & 231308507 & [CW83] 0512$-$08 & 2\\
	6056031847366325760 & 437002682 & SALT J125657.1$-$601950 & 3 & 5762960596350664832 & 280775141 & [CW83] 0904$-$02 & 2\\
	5843886507606067200 & 338653597 & SALT J131607.5$-$701236 & 2 & & & & \\
    \hline
    \end{tabular}
    \caption{Table \ref{tab:app2} cont.}
    \label{tab:app3}
\end{table*}

%%%%%%%%%%%%%%%%%%%%%%%%%%%%%%%%%%%%%%%%%%%%%%%%%%

% Don't change these lines
\bsp	% typesetting comment
\label{lastpage}
\end{document}